\documentclass[pra, twocolumn,notitlepage,nofootinbib,superscriptaddress]{revtex4-1}
\usepackage{pifont}%
\usepackage{bbold}%
\usepackage{braket}
\usepackage[dvipsnames]{xcolor} %
\usepackage{amsmath, amssymb, graphicx, bm}
\usepackage{slashed}
\usepackage{subcaption}
\usepackage{comment}
\usepackage{pifont}%

\usepackage{algorithm}
\usepackage{algpseudocode}
\DeclareMathOperator{\sign}{sign}
\usepackage[framemethod=TikZ]{mdframed}
\newenvironment{Frame}[1][]{%
    \begin{mdframed}[%
   	 frametitle={#1},
   	 skipabove=\baselineskip plus 2pt minus 1pt,
   	 skipbelow=\baselineskip plus 2pt minus 1pt,
   	 linewidth=0.5pt,
   	 frametitlerule=true,
   	 frametitlebackgroundcolor=gray!30,
   	 roundcorner=10pt,
   	 nobreak=true
   	 ]%
    }{%
    \end{mdframed}
}

\begin{document}
    
    \title{Matrix decompositions in Quantum Optics: \\ Takagi/Autonne, Bloch-Messiah/Euler, Iwasawa, and Williamson}

    \author{Martin Houde}
    \email{martin.houde@polymtl.ca}
 \affiliation{D\'epartement de g\'enie physique, \'Ecole Polytechnique de Montr\'eal, Montr\'eal, QC, H3T 1J4, Canada}
 
 \author{Will McCutcheon}
 \email{W.McCutcheon@hw.ac.uk}
 \affiliation{Institute of Photonics and Quantum Sciences, Heriot-Watt University, Edinburgh EH14 4AS, UK}
 
	 \author{Nicol\'as Quesada}
    \email{nicolas.quesada@polymtl.ca}
 \affiliation{D\'epartement de g\'enie physique, \'Ecole Polytechnique de Montr\'eal, Montr\'eal, QC, H3T 1J4, Canada}
    \date{\today}
 \begin{abstract}
 
In this note we summarize four important matrix decompositions commonly used in quantum optics, namely the Takagi/Autonne, Bloch-Messiah/Euler, Iwasawa, and Williamson decompositions. The first two of these decompositions are specialized versions of the singular-value decomposition when applied to symmetric or symplectic matrices. The third factors any symplectic matrix in a unique way in terms of matrices that belong to different subgroups of the symplectic group. 
The last one instead gives the symplectic diagonalization of real, positive definite matrices of even size. While proofs of the existence of these decompositions exist in the literature, we focus on providing explicit constructions to implement these decompositions using standard linear algebra packages and functionalities such as singular-value, polar, Schur and QR decompositions, and matrix square roots and inverses.
 \end{abstract}

\maketitle

\section{Introduction}
In this note we provide explicit constructions to obtain several commonly used matrix decompositions in Quantum Optics which can be readily implemented using standard numerical libraries and functions. Each of the decompositions described here has been implemented and tested in the module \texttt{decompositions} of \texttt{thewalrus}~\cite{gupt2019walrus} in \texttt{Python} using standard functions from the \texttt{numpy}~\cite{harris2020array} and \texttt{scipy}~\cite{2020SciPy-NMeth} numerical linear algebra packages and also in the \texttt{SymplecticDecompositions.jl} package~\cite{quesadasymplectic} written in \texttt{Julia}~\cite{bezanson2017julia}.

In Sec.~\ref{sec:LA} we provide some basic review of linear algebra results, including normal matrices and matrix square roots, as well as eigen, polar, singular-value, Schur and QR decompositions. References to literature where the proofs of the existence of these decompositions are provided. These decompositions have already been implemented in \texttt{LAPACK}~\cite{lapack99} and thus are available in most (if not all) common numerical linear algebra packages (including the aforementioned \texttt{scipy}, \texttt{numpy} and \texttt{Julia}).

In Sec.~\ref{sec:symplectic} we discuss some basic facts about the symplectic group which plays an important role later on. The discussion presented borrows from many excellent references such as Serafini~\cite{serafini2017quantum}, Arvind et al.~\cite{arvind1995real} and Nicacio~\cite{nicacio2021williamson}.

In Sec.~\ref{sec:takagi} we discuss the Takagi/Autonne decomposition of a square real- or complex-symmetric matrix, which is a specialized singular-value decomposition that takes advantage of the symmetry of the input matrix. As noted in page 159 of Horn and Johnson~\cite{horn2012matrix} this decomposition was introduced by L. Autonne in 1915. According to Ref.~\cite{brezinski2022journey} it was then rediscovered by T. Takagi in 1925~\cite{brezinski2022journey}. This decomposition finds applications when introducing Schmidt modes of biphotons~\cite{fabre2020modes} or squeezed states~\cite{arzani2018versatile,quesada2022beyond} and also when encoding matrices into Gaussian Boson Samplers~\cite{jahangiri2020point,oh2023quantum}, which are subuniversal photonic quantum computers~\cite{hamilton2017gaussian,kruse2019detailed}.

In Sec.~\ref{sec:bm} we discuss the Bloch-Messiah/Euler decomposition, which as Serafini~\cite{serafini2017quantum} notes is ``nothing but the singular
value decomposition of a symplectic matrix''. The name Bloch-Messiah was introduced by Braunstein~\cite{braunstein2005squeezing} as this is the Boson ``formal
extension of the original result for fermions'' derived by C. Bloch and A. Messiah in Ref.~\cite{bloch1962canonical}.  The Euler name comes from ``a faint analogy with the Euler decomposition
of orthogonal transformations''~\cite{serafini2017quantum}.
This decomposition finds application in e.g. identifying irreducible resources in continuous-variable (CV) quantum information~\cite{braunstein2005squeezing,serafini2007standard}, developing compilers for CV quantum computers~\cite{kalajdzievski2021exact,cariolaro2016bloch,cariolaro2016reexamination}, and identifying input- and output modes of optical squeezers~\cite{fabre2020modes,houde2024perfect,jiang2012time}.

In Sec.~\ref{sec:iwasawa} we discuss the pre-Iwasawa and Iwasawa decompositions. In Ref.~\cite{iwasawa1949some}, K. Iwasawa introduced the decomposition that bears his name for general semisimple Lie groups. We provide details of this decomposition for the specific case of the symplectic group. 
Both of these find applications in the study of group-theoretic properties of the symplectic group (see for example Arvind~\cite{arvind1995real},  Prop. 2.29 of Gosson~\cite{de2006symplectic}, Page 2 of Haberman and 
Haberman~\cite{habermann2006introduction} and Page 179 of Folland~\cite{folland1989harmonic}). Moreover, the Iwasawa decomposition finds applications in e.g.,  obtaining keyrate bounds in quantum key distribution~\cite{lodewyck2007tight} and discussing properties of CV cluster states~\cite{gonzalez2021cluster}.

Finally, in Sec.~\ref{sec:will} we discuss the Williamson decomposition which can be used to diagonalize positive-definite matrices of even size using a symplectic matrix. This decomposition is a corollary of the general results derived by J. Williamson in Ref.~\cite{williamson1936algebraic} (see also Appendix 6 of Arnol'd~\cite{arnol2013mathematical} and Ref.~\cite{son2022symplectic} for possible generalizations to positive semi-definite matrices).  This decomposition finds applications in many areas of CV quantum information such as the calculation of entanglement measures of Gaussian CV states~\cite{weedbrook2012gaussian, adesso2007entanglement}, effects of filtering in twin-beams~\cite{houde2023waveguided}, pure state decompositions of mixed Gaussian states~\cite{weedbrook2012gaussian}, fidelity of Gaussian states~\cite{banchi2015quantum}, and in speeding the classical simulation of gaussian boson sampling~\cite{quesada2022quadratic}.

\section{Linear algebra results}\label{sec:LA}
In this section, we review some basic concepts in linear algebra; for a thorough discussion cf. e.g. Horn and Johnson~\cite{horn2012matrix}.
We use boldface letters to denote a matrix such as $\bm{A}$ with $(i,j)$ entry $A_{i,j}$. We use $\bm{A}^*, \bm{A}^T$ and $\bm{A}^\dagger$ to denote the conjugate, transpose and conjugate-transpose of the matrix $\bm{A}$ with entries $A_{i,j}^*$, $A_{j,i}$ and $A_{j,i}^*$ respectively. In this manuscript we are only interested in square matrices with complex or real entries.

\subsection{The spectral theorem and matrix square roots}
We recall that a matrix $\bm{A}$ is normal if it commutes with its conjugate transpose $\bm{A}^\dagger$, i.e., $\bm{A} \bm{A}^\dagger - \bm{A}^\dagger \bm{A}=0$.
The spectral theorem states that a matrix $\bm{A}$ is unitarily diagonalizable if and only if $\bm{A}$ is normal (cf. Theorem 2.5.3. from Horn and Johnson~\cite{horn2012matrix}).
A unitary diagonalization (``eigendecomposition'') of a matrix $\bm{A}$ allows one to write it as
\begin{align}\label{eq:diagonalized}
    \bm{A} = \bm{U} \left[\oplus_{i=1}^{\ell} a_{i} \right] \bm{U}^\dagger,
\end{align}
where $\bm{U}$ is a unitary matrix, $\ell$ is the dimension of the square matrix $\bm{A}$, and $a_i$ are the eigenvalues of $\bm{A}$. We use direct sum notation for scalars such that $\oplus_{i=1}^{\ell} a_{i}~=~\text{diag}(a_{1},a_{2},\ldots,a_{\ell})$ is a diagonal matrix. 

Important instances of normal matrices are hermitian matrices for which $\bm{A} = \bm{A}^\dagger$ and unitary matrices for which $\bm{W} \bm{W}^\dagger =  \bm{W}^\dagger \bm{W} = \mathbb{1}_\ell$; moreover for these two types of matrices the eigenvalues are real and of unit complex absolute value respectively.
Note that for unitarily diagonalizable matrices, one can introduce matrix square roots. For a matrix unitarily diagonalized as in Eq.~\eqref{eq:diagonalized} one introduces matrix square roots as follows
\begin{align}
\sqrt{\bm{A}} \equiv \bm{U} \left[\oplus_{i=1}^{\ell} ( \pm \sqrt{a_{i}}) \right] \bm{U}^\dagger.
\end{align}
Note that we use the term \emph{a} matrix square root and not \emph{the} matrix square root since one can pick the plus or minus sign for any of the eigenvalues appearing in the last equation. Regardless of the choice of sign, note that any matrix square root satisfies $\sqrt{\bm{A}} \sqrt{\bm{A}} = \bm{A}$.
In the special case where all the eigenvalues are non-negative (positive), the unitarily diagonalizable matrix is termed positive semi-definite PSD (or positive definite PD) and then its unique PSD (PD) square root is obtained by always taking the plus signs in the equation above.
The spectral theorem also applies to real-symmetric matrices where in Eq.~\eqref{eq:diagonalized} the unitaries are replaced by orthogonal matrices and $a_i \in \mathbb{R}$. Similarly, one can construct real PSD matrices as the symmetric matrices that have non-negative eigenvalues, these too have unique square roots, just like in the hermitian case. These conclusions, namely the replacement of unitary matrices by orthogonal matrices when discussing real-symmetric instead complex-hermitian matrices extend to the polar and singular-value decompositions discussed below.

\subsection{The polar decomposition}\label{sec:polar}
Given a square matrix $\bm{A}$ the polar decomposition states that one can factorize it as the product of a PSD matrix $\bm{P}$ and a unitary matrix $\bm{W}$,
\begin{align}\label{eq:polardef}
\bm{A} = \bm{P} \bm{W},
\end{align}
where $\bm{P} = \sqrt{\bm{A} \bm{A}^\dagger}$.  If $\bm{A}$ is invertible then $\bm{W} = \bm{P}^{-1} \bm{A}$. For a proof see Caves~\cite{caves} or Theorem 7.3.1. of Horn and Johnson~\cite{horn2012matrix}.
An important property of normal matrices is that the $\bm{P}$ and $\bm{W}$ in their polar decomposition commute. Moreover,  $\sqrt{\bm{P}}$ and $\bm{W}$ also commute.
To see why this is the case note that if $\bm{A}$ is normal then
\begin{align}\label{eq:polarcomm}
\bm{P}^2 = \bm{A} \bm{A}^\dagger = \bm{A}^\dagger \bm{A} = \bm{W}^\dagger \bm{P}^2 \bm{W}.
\end{align}
Since the matrices on the left-hand side and right-hand side are PSD we can take square roots to find
\begin{align}
\bm{P} =  \bm{W}^\dagger \bm{P} \bm{W} \longleftrightarrow \bm{W} \bm{P} = \bm{P} \bm{W}.
\end{align}
If we take the square root twice in Eq.~\eqref{eq:polarcomm}, we easily find that $[\sqrt{\bm{P}}, \bm{W}] = 0$ for normal matrices.

\subsection{Singular-value decomposition}
Once the polar decomposition of a matrix is at hand, one can further build a singular-value decomposition (SVD) by unitarily diagonalizing the PSD matrix $\bm{P}$ in Eq.~\eqref{eq:polardef} as
\begin{align}
\bm{P} = \bm{U} \left[\oplus_{i=1}^{\ell} a_{i} \right] \bm{U}^\dagger,
\end{align}
plugging back in to the polar decomposition of $\bm{A}$ we find
\begin{align}
\bm{A} = \bm{U} \left[\oplus_{i=1}^{\ell} a_{i} \right] \bm{U}^\dagger \bm{W}.
\end{align}
Since both $\bm{U}^\dagger$ and $\bm{W}$ are unitary, their product is unitary and thus we can identify a unitary matrix (or its transpose, conjugate or conjugate-transpose) as $\bm{V}^\dagger =  \bm{U}^\dagger \bm{W}$ ($\bm{V}^* =  \bm{U}^\dagger \bm{W}$, $\bm{V}^T =  \bm{U}^\dagger \bm{W}$ or $\bm{V} =  \bm{U}^\dagger \bm{W}$ respectively) to write the SVD of $\bm{A}$ as
\begin{align}
\bm{A} = \bm{U} \left[\oplus_{i=1}^{\ell} a_{i} \right] \bm{V}^\dagger.
\end{align}
Note that the singular values $a_i$ are all non-negative, as they are the eigenvalues of a PSD matrix.

\subsection{The real Schur decomposition}\label{sec:schur}

From theorem 7.4.1 of Golub and van Loan~\cite{golub2013matrix} we have that every real square matrix $\bm{A}$ of size $\ell$ can be written as
\begin{align}
    \bm{A} = \bm{O} \bm{R}  \bm{O}^T,
\end{align}
where $\bm{O}$ is an orthogonal matrix and
\begin{align}\label{eq:schur}
    \bm{R} = \begin{pmatrix}
   	 \bm{R}_{11} &\bm{R}_{12} &\ldots & \bm{R}_{1m} \\
   	 0 & \bm{R}_{22} & \ldots & \bm{R}_{2m} \\
   	 \vdots & \vdots & \ddots & \vdots \\
   	 0 & 0 & \ldots & \bm{R}_{mm}
    \end{pmatrix},
\end{align}
where each $\bm{R}_{ii}$ is either a 1-by-1 matrix or a 2-by-2 matrix having complex-conjugate eigenvalues.

Note that if $\bm{A}$ is antisymmetric and of even dimensions (as it is below) $\bm{A} = -\bm{A}^T$ then, one can apply the spectral theorem to the hermitian matrix $i \bm{A}$ to show that the eigenvalues of $\bm{A}$ are purely imaginary. Since the matrix $\bm{A}$ is real, then, they come in complex conjugate pairs of the form $\pm i a, a\in \mathbb{R}$.
Relative to Eq.~\eqref{eq:schur} and bearing the conclusion from the last paragraph it is easy to see that for an antisymmetric $\bm{A}$ it must be that $\bm{R}_{ij} = 0$ if $i \neq j$ and that each block diagonal matrix must have the form $\bm{R}_{ii} =  \left(\begin{smallmatrix} 0 &\pm {a_i}  \\ \mp{a_i} & 0 \end{smallmatrix} \right)$ with $a_i \geq 0$. This observation giving a quasi-diagonalization for real antisymmetric matrices can also be derived in a more self-contained way without invoking the Schur decomposition as done by C. Caves in Ref.~\cite{caves_anti}.

\subsection{The QR decomposition}
For square real matrices, $\bm{A} \in \mathbb{R}^{\ell \times \ell}$, the QR decomposition states that any matrix from this set can be factored as
\begin{align}
	\bm{A} = \bm{Q} \bm{R} 	
\end{align}
where $\bm{Q}$ is orthogonal, $\bm{Q} \bm{Q}^T = \mathbb{1}_{\ell}$, and $\bm{R}$ is upper triangular (i.e. $R_{i,j} = 0$ if $i>j$). A proof of the existence of this decomposition is given in Sec. 5.2. of Golub and van Loan~\cite{golub2013matrix}. In this same reference it is shown that if the matrix $\bm{A}$ is invertible then this decomposition is unique if the diagonal elements of $\bm{R}$ are chosen to be positive.

\section{The Symplectic Group}\label{sec:symplectic}

The real symplectic group~\cite{arvind1995real,serafini2017quantum} is defined as the set of matrices of even size $2\ell$ that satisfy
\begin{align}\label{eq:sympcond}
    \mathrm{Sp}(2\ell, \mathbb{R}) = \{\bm{S}\in \mathbb{R}^{2\ell\times 2\ell}| \bm{S\Omega S}^T = \bm{\Omega}\},
\end{align}
where the symplectic form is given by,
\begin{align}
    \bm{\Omega} = \begin{pmatrix}
   	 0_{\ell} & \mathbb{1}_{\ell} \\
    -\mathbb{1}_{\ell} & 0_{\ell}     
    \end{pmatrix}.\label{eq:sympform}
\end{align}
If we partition an even size matrix in equal sized blocks
\begin{align}
	\bm{S} = \begin{pmatrix}
		\bm{A} & \bm{B} \\\bm{C} & \bm{D}
	\end{pmatrix},
\end{align}
then the conditions for being an element of the symplectic group translate into the following for the blocks
\begin{subequations}\label{eq:sympblocks}
\begin{align}
\bm{A}^T \bm{C} = \bm{C}^T \bm{A}, \ \bm{B}^T \bm{D} = \bm{D}^T \bm{C}, \ \bm{A}^T \bm{D} - \bm{C}^T \bm{B} = \mathbb{1}_\ell, \\
\bm{A} \bm{B}^T = \bm{B} \bm{A}^T, \ \bm{C} \bm{D}^T = \bm{D} \bm{C}^T, \ \bm{A} \bm{D}^T - \bm{B} \bm{C}^T = \mathbb{1}_\ell. 
\end{align}
\end{subequations}
This group has several interesting properties which are useful in certain derivations. The symplectic form, $\bm{\Omega}$, has the following properties
\begin{align}
    \bm{\Omega} \in \mathrm{Sp}(2\ell,\mathbb{R}),\\
    \bm{\Omega}^{-1} = \bm{\Omega}^T = - \bm{\Omega} \in \mathrm{Sp}(2\ell,\mathbb{R}).
\end{align}
By taking the determinant of the symplectic condition in Eq.~\eqref{eq:sympcond} one easily finds
\begin{align}\label{eq:det}
	|\det(\bm{S})| = 1.
\end{align}
Therefore, all symplectic matrices are invertible and moreover
\begin{align}\label{eq:Sinverse}
	\bm{S}^{-1} = \bm{\Omega} \bm{S}^{T} \bm{\Omega}^{T} \in \mathrm{Sp}(2\ell,\mathbb{R}).
\end{align}
From this we can also show that if $\bm{S}\in\mathrm{Sp}(2\ell,\mathbb{R})$ then so is its transpose, $\bm{S}^{T}\in\mathrm{Sp}(2\ell,\mathbb{R})$. Since all symplectic matrices are invertible, they can all be uniquely decomposed into polar form
\begin{align}
	\bm{S} = \bm{P}\bm{O},
\end{align}
where by construction $\bm{P}=\bm{P}^{T}$ is PD and $\bm{O}$ is orthogonal. Since this decomposition is unique, we can also show the both $\bm{P}\in\mathrm{Sp}(2\ell,\mathbb{R})$ and $\bm{O}\in\mathrm{Sp}(2\ell,\mathbb{R})$.

To see this, we consider the symplectic definition
\begin{align}
	\bm{S\Omega S}^T = \bm{\Omega}
	\implies \bm{S} &= \bm{\Omega} \left( \bm{S}^{T} \right)^{-1} \bm{\Omega}^{T}\nonumber\\
                	&=\bm{\Omega} \left( \bm{O}^{T}\bm{P} \right)^{-1} \bm{\Omega}^{T}\nonumber\\
                	&=\bm{\Omega} \bm{P}^{-1} \bm{O} \bm{\Omega}^{T}\nonumber\\
                	&=\left(\bm{\Omega}  \bm{P}^{-1}\bm{\Omega}^{T}\right) \left(\bm{\Omega} \bm{O} \bm{\Omega}^{T}\right).
\end{align}
Since $\bm{P}$ is symmetric PD then so is $\bm{P}^{-1}$. The product $\left(\bm{\Omega}  \bm{P}^{-1}\bm{\Omega}^{T}\right)$ is symmetric and also PD as the symplectic matrix is orthogonal.

The second product, $\left(\bm{\Omega} \bm{O} \bm{\Omega}^{T}\right)$, is orthogonal. We have thus obtained a second expression for the polar decomposition of $\bm{S}$ and by uniqueness
\begin{align}
	\bm{P} = \bm{\Omega}  \bm{P}^{-1}\bm{\Omega}^{T} &\implies \bm{P}\bm{\Omega}\bm{P}=\bm{\Omega},\\
	\bm{O} = \bm{\Omega}  \bm{O}\bm{\Omega}^{T} &\implies \bm{O}\bm{\Omega}\bm{O}^{T}=\bm{\Omega},
\end{align}
which proves that $\bm{P}\in\mathrm{Sp}(2\ell,\mathbb{R})$ and $\bm{O}\in\mathrm{Sp}(2\ell,\mathbb{R})$. This uniqueness can be used to show that the absolute value can be removed in Eq.~\eqref{eq:det} without loss of generality for any symplectic matrix.

\subsection{Complex Form}

We define the complex form of $\mathrm{Sp}(2\ell, \mathbb{R})$ denoted by $\mathrm{Sp_{c}}(2\ell)$ as
\begin{align}
	\mathrm{Sp_{c}}(2\ell) = \{ \bm{\mathcal{S}}\in\mathbb{C}^{2\ell\times2\ell}| \bm{\mathcal{S}}=\bm{R}^{\dagger}\bm{S}\bm{R}; \bm{S}\in\mathrm{Sp}(2\ell,\mathbb{R})	\},
\end{align}
where
\begin{align}
	\bm{R} = \frac{1}{\sqrt{2}}\begin{pmatrix}
    	\mathbb{1}_{\ell} & \mathbb{1}_{\ell} \\
    	-i\mathbb{1}_{\ell} & i\mathbb{1}_{\ell}\\
	\end{pmatrix}
\end{align}
is a unitary matrix. Note that this is not the same as $\mathrm{Sp}(2\ell,\mathbb{C})$, which is the set of complex matrices that are symplectic. The matrices $\bm{\mathcal{S}}$ have interesting properties as well. By applying the transformation to the symplectic definition, we find that matrices $\bm{\mathcal{S}}\in\mathrm{Sp_{c}}(2\ell)$ obey
\begin{align}\label{eq:complexsympform}
	\bm{\mathcal{S}}\bm{Z}\bm{\mathcal{S}}^{\dagger} = \bm{Z},
\end{align}
where
\begin{align}
	\bm{Z} = \begin{pmatrix}
    	\mathbb{1}_{\ell} & 0_{\ell} \\
    	0_{\ell} & -\mathbb{1}_{\ell}\\
	\end{pmatrix}.
\end{align}
Furthermore, by expressing $\bm{S}\in\mathrm{Sp}(2\ell,\mathbb{R})$ in block form and applying the transformation, we find that the matrices $\bm{\mathcal{S}}$ take a very specific form
\begin{align}\label{eq:complexform}
	\bm{\mathcal{S}}=\begin{pmatrix}
    	\bm{H} & \bm{K} \\
    	\bm{K}^{*} & \bm{H}^{*}\\
	\end{pmatrix},
\end{align}
for some $\ell \times \ell$ blocks $\bm{H}$ and $\bm{K}$. We can obtain further constraints on the block matrix form of $\bm{\mathcal{S}}$ given extra conditions on the matrices $\bm{S}$. Consider an orthogonal symplectic matrix $\bm{O}$ and its complex form $\bm{\mathcal{U}}=\bm{R}^{\dagger}\bm{O}\bm{R}$. It is straightforward to see that $\bm{\mathcal{U}}$ is unitary. To satisfy unitarity, the condition of Eq.~\eqref{eq:complexsympform}, and the condition of Eq.~\eqref{eq:complexform} the block matrix structure of $\bm{\mathcal{U}}$ must be
\begin{align}\label{eq:Ublock}
	\bm{\mathcal{U}}=\begin{pmatrix}
    	\bm{U} & 0_{\ell} \\
    	0_{\ell} & \bm{U}^{*}
	\end{pmatrix} = \exp\left[ i
 \begin{pmatrix}
    	\bm{J} & 0_{\ell} \\
    	0_{\ell} & -\bm{J}^T
	\end{pmatrix}
 \right],
\end{align}
where the block $\bm{U} = \exp(i \bm{J})$ is unitary and $\bm{J} = \bm{J}^\dagger$ is its hermitian generator. If we consider a symmetric PD symplectic matrix $\bm{P}$ and its complex form $\bm{\mathcal{P}} = \bm{R}^{\dagger}\bm{P}\bm{R}$, we see that $\bm{\mathcal{P}}=\bm{\mathcal{P}}^{\dagger}$ is hermitian. Furthermore, the characteristic polynomial of $\bm{\mathcal{P}}$ is equivalent to that of $\bm{P}$ and so $\bm{\mathcal{P}}$ is also PD. To satisfy hermiticity and the conditions of Eqs.~\eqref{eq:complexsympform},~\eqref{eq:complexform} we find that the block matrix form is, similarly to $\bm{\mathcal{S}}$,
\begin{align}\label{eq:Hermitionblock}
	\bm{\mathcal{P}}=\begin{pmatrix}
    	\bm{N} & \bm{M} \\
    	\bm{M}^{*} & \bm{N}^{*}
	\end{pmatrix},
\end{align}
but now $\bm{M}=\bm{M}^{T}$ is symmetric and $\bm{N}$ is hermitian. Since $\bm{N}$ is the diagonal block of a PD matrix, it too is PD. With these constraints we can consider the polar decomposition of matrices $\bm{\mathcal{S}}\in\mathrm{Sp_{c}}(2\ell)$. From the definition
\begin{align}\label{eq:complexpolar}
	\bm{\mathcal{S}} &= \bm{R}^{\dagger}\bm{S}\bm{R} \nonumber\\
    	&=\bm{R}^{\dagger}\bm{PO}\bm{R}\nonumber\\
    	&=\bm{R}^{\dagger}\bm{P}\bm{R}\bm{R}^{\dagger}\bm{O}\bm{R}\nonumber\\
    	&=\bm{\mathcal{P}}\bm{\mathcal{U}},
\end{align}
where $\bm{\mathcal{U}}$ is unitary with the block structure of Eq.~\eqref{eq:Ublock} and $\bm{\mathcal{P}}$ is hermitian PD with the block structure of Eq.~\eqref{eq:Hermitionblock}. This tells us that Eq.~\eqref{eq:complexpolar} is the unique polar decomposition of the complex forms $\bm{\mathcal{S}}$. The properties above are important when considering the Bloch-Messiah/Euler decomposition of real symplectic matrices.

\section{Takagi/Autonne decomposition}\label{sec:takagi}
Given a symmetric (and in general complex) matrix $\bm{M} = \bm{M}^T \in \mathbb{C}^{\ell \times \ell}$ its Takagi/Autonne decomposition is given by
\begin{align}\label{def:takagi}
\bm{M} = \bm{W} \bm{\Lambda} \bm{W}^T,\quad  \bm{\Lambda} =  \oplus_{i=1}^\ell \lambda_i,
\end{align}
where $\bm{W}$ is unitary, $\bm{W}\bm{W}^\dagger = \mathbb{1}_\ell$ and $\lambda_i \geq 0$. This decomposition is an SVD, albeit one with a special symmetry that makes it explicit that the matrix being decomposed is symmetric; indeed the decomposition on the right-hand side of the Eq.~\eqref{def:takagi} makes it explicit that the object on the left-hand side is symmetric.

To obtain the decomposition we first obtain an SVD of $\bm{M}$
\begin{align}\label{svd}
\bm{M} = \bm{U} \bm{\Lambda} \bm{V}^\dagger.
\end{align}
For arbitrary matrices there is nothing special we can say about the relation between the unitary matrices $\bm{U}$ and $\bm{V}$. However, $\bm{M}$ is no arbitrary matrix and indeed we will show that for \emph{symmetric} $\bm{M}$ the product
\begin{align}
\bm{V}^T \bm{U}
\end{align}
has a number of special properties that will allow us to construct the sought-after $\bm{W}$ in Eq.~\eqref{def:takagi}. To this end let us introduce the following matrix
\begin{align}
\bm{L} = \bm{V}^T \bm{M} \bm{V}.
\end{align}
Since $\bm{M}=\bm{M}^T$ then it holds that $\bm{L}=\bm{L}^T$ is symmetric. Moreover when using the SVD of $\bm{M}$ in Eq.~\eqref{svd} we obtain the polar decomposition
\begin{align}
\bm{L} = \bm{V}^T \left(\bm{U} \bm{\Lambda} \bm{V}^\dagger \right) \bm{V} = \underbrace{\bm{V}^T \bm{U}}_{\text{Unitary}} \cdot  \underbrace{\bm{\Lambda}}_{\text{PSD}}.
\end{align}
It is also easy to verify that the matrix $\bm{L}$ is normal as
\begin{align}
    \bm{L}^\dagger \bm{L} &= \bm{V}^\dagger \bm{M}^\dagger \bm{M} \bm{V} =  \bm{\Lambda}^2\\
    \bm{L}\bm{L}^\dagger &=  \bm{V}^T \bm{M} \bm{V} \bm{V}^\dagger \bm{M}^\dagger \bm{V}^* = \bm{V}^T \bm{M} \bm{M} ^\dagger \bm{V}^* \\
    &= (\bm{V}^\dagger \bm{M}^\dagger \bm{M}  \bm{V})^* =	\bm{\Lambda}^{*2} =  \bm{\Lambda}^2.
\end{align}
Using the results from Sec.~\ref{sec:polar}, we thus have
\begin{align}
    [\bm{V}^T \bm{U}, \bm{\Lambda}]=    [\bm{V}^T \bm{U}, \sqrt{\bm{\Lambda}}]=0,
\end{align}
which allows us to write
\begin{align}
    \bm{L} = \bm{V}^T \bm{U} \sqrt{\bm{\Lambda}} \sqrt{\bm{\Lambda}} = \sqrt{\bm{\Lambda}} \bm{V}^T\bm{U} \sqrt{\bm{\Lambda}} .
\end{align}
Note that if $\bm{\Lambda}$ has all distinct singular values, then it automatically holds that $\bm{V}^T \bm{U}$ is also diagonal as there are no degenerate subspaces. In the more general case where there are degeneracies, it must be that $\bm{V}^T \bm{U}$ is block-diagonal with each block corresponding to a degenerate subspace labelled by a degenerate singular value; the dimension of each degenerate block is precisely the number of times a given singular values is repeated in $\bm{\Lambda}$. Note moreover, that within each subspace labelled by $\lambda_i$ the matrix $\bm{\Lambda}$ acts as a multiple of the identity matrix, this immediately implies that
\begin{align}\label{eq:commute}
    [\sqrt{\bm{V}^T \bm{U}}, \bm{\Lambda}]=0.
\end{align}
These equations allow us to conclude that in any block of the matrix $\bm{V}^T\bm{U}$ that does not correspond to null singular values, the matrix  $\bm{V}^T\bm{U}$ must be symmetric. Formally, if we assume without loss of generality that the singular values are ordered in decreasing order and that there are $k$ that are zero, and introducing the projection matrix
\begin{align}
\bm{K} = \mathbb{1}_{\ell -k} \oplus 0_k,
\end{align}
then it holds that
\begin{align}\label{def:Q}
\bm{Q} \equiv  \bm{K} \bm{V}^T \bm{U} \bm{K} =  \bm{V}^T \bm{U} \bm{K} = \bm{K} \bm{V}^T \bm{U},
\end{align}
is symmetric, $\bm{Q} = \bm{Q}^T$. Moreover, by construction this projector satisfies
\begin{align}\label{def:LambdaQ}
\bm{\Lambda} \bm{K} = \bm{K}\bm{\Lambda} = \bm{K} \bm{\Lambda} \bm{K}  = \bm{\Lambda}.
\end{align}
Putting Eq.~\eqref{def:Q} and Eq.~\eqref{def:LambdaQ} together one easily obtains
\begin{align}\label{eq:lambdaQeq}
\bm{\Lambda} \bm{V}^T \bm{U} =  \bm{V}^T \bm{U} \bm{\Lambda} =  \bm{\Lambda} \bm{Q} = \bm{Q} \bm{\Lambda}.
\end{align}

In summary, relative to the SVD of the symmetric matrix $\bm{M}$ in Eq.~\eqref{svd} we have that the block-diagonal unitary matrices $\bm{V}^T\bm{U}$ and $\sqrt{\bm{V}^T\bm{U}}$ commute with $\bm{\Lambda}$ and moreover, the matrix $\bm{Q}$ in Eq.~\eqref{def:Q} is symmetric.

With these lemmas proven, we now claim that the sought-after Takagi/Autonne unitary is
\begin{align}
    \bm{W} = \bm{U} \sqrt{(\bm{U}^T \bm{V})^*},
\end{align}
which we confirm straightforwardly using the above derived properties of  $\bm{\Lambda}$, $\bm{V}^T\bm{U}$ and $\sqrt{\bm{V}^T\bm{U}}$ as follows
\begin{subequations}\label{eq:proof}
\begin{align}
\bm{W} \bm{\Lambda} \bm{W}^T =& \bm{U} \sqrt{(\bm{U}^T \bm{V})^*} \bm{\Lambda} \sqrt{(\bm{U}^T \bm{V})^{\dagger}}\bm{U} ^T\\ \label{eq:proof1}
=& \bm{U} \bm{\Lambda} \sqrt{(\bm{K}\bm{U}^T \bm{V})^*}  \sqrt{(\bm{K}\bm{U}^T \bm{V})^{\dagger}}\bm{U} ^T  \\ \label{eq:proof4}
=& \bm{U} \bm{\Lambda} \sqrt{\bm{Q}^*} \sqrt{\bm{Q}^*}\bm{U} ^T  \\ \label{eq:proof6}
=& \bm{U} \bm{\Lambda} \bm{V}^{\dagger} \bm{U}^* \bm{U} ^T   \\ \label{eq:proof8}
=& \bm{U} \bm{\Lambda} \bm{V}^\dagger  = \bm{M}.
\end{align}
\end{subequations}
In Eq.~\eqref{eq:proof1} we have used the commutative property of Eq.~\eqref{eq:commute} as well as  Eq.~\eqref{def:LambdaQ} and Eq.~\eqref{def:Q} to introduce the projection matrices $\bm{K}$ which we can and have moved it into the square-roots. In Eq.~\eqref{eq:proof4} we have used the definition of $\bm{Q}$ along with the fact that it is symmetric. Finally, in Eq.\eqref{eq:proof8} we have used Eq.~\eqref{eq:lambdaQeq}. This derivation partially follows the presentation of Caves~\cite{caves} and also the results in Ref.~\cite{chebotarev2014singular} (where it is shown to be numerically stable) and is summarized in the box below.

\begin{Frame}[Box 1. Takagi/Autonne decomposition]
    Input: Symmetric matrix $\bm{M} = \bm{M}^T$.\\
    Calculate the SVD $\bm{M} = \bm{U} \left[ \oplus_{i=1}^\ell \lambda_i \right] \bm{V}^\dagger$.\\
    Return:      $ \bm{\Lambda} = \oplus_{i=1}^\ell \lambda_i$, 
    $\bm{W} =  \bm{U} \sqrt{(\bm{U}^T \bm{V})^*}$.
\end{Frame}

Consider the special case where $\bm{M}$ is \emph{real} and symmetric. We can immediately write an eigendecomposition as
\begin{align}
\bm{M} = \bm{O} \left[  \oplus_{i=1}^\ell r_i  \right] \bm{O}^T,
\end{align}
where $\bm{O}$ is real and orthogonal (satisfying $\bm{O} \bm{O}^T = \mathbb{1}_\ell$) and $r_i \in \mathbb{R}$ are the eigenvalues of $\bm{M}$. We can, from the above eigendecomposition, obtain the Takagi/Autonne decomposition by writing $r_i = |r_i| \sign r_i = \sqrt{\sign r_i} |r_i| \sqrt{\sign r_i}$ where we define $\sign x = +1 $ if $x \geq 0$ and $-1$ if $x<0$. With this notation we have
\begin{align}
\bm{M} =& \bm{O} \left[ \oplus_{i=1}^\ell |r_i| \sign r_i  \right] \bm{O}^T \\
=& \underbrace{\bm{O} \left[ \oplus_{i=1}^\ell \sqrt{\sign r_i} \right]}_{\equiv \bm{W}} \underbrace{\left[ \oplus_{i=1}^\ell |r_i| \right]}_{\equiv \bm{\Lambda}}   \underbrace{\left[ \oplus_{i=1}^\ell \sqrt{\sign r_i} \right] \bm{O}^T. }_{\equiv \bm{W}^T}
\end{align}
Note that the matrix $\bm{O} \left[ \oplus_{i=1}^\ell \sqrt{\sign r_i} \right]$ is unitary as it is the product of an orthogonal matrix and a diagonal unitary matrix. Note that if $\bm{M}$ is real, symmetric and PSD (i.e. $r_i \geq 0$ for all $i$) then the Takagi/Autonne decomposition and its eigendecomposition coincide.

\section{Bloch-Messiah/Euler Decomposition}\label{sec:bm}

A real symplectic matrix $\bm{S}$ can be decomposed as     
\begin{align}
    \bm{S} = \bm{O} \bm{D} \bm{Q},
\end{align}
where
\begin{align}
    \bm{D} = \bm{\Gamma} \oplus \bm{\Gamma}^{-1},\quad \bm{\Gamma} = \oplus_{i=1}^\ell \gamma_i,
\end{align}
and $\gamma_i \geq  1$. $\bm{O},\bm{Q}\in \mathrm{C}(\ell)$ where $\mathrm{C}(\ell)$ denotes the maximal compact subgroup of the symplectic which is isomorphic to the unitary group of size $\ell$, $\mathrm{C}(\ell)~=~\mathrm{Sp}(2\ell,\mathbb{R}) \cap \mathrm{O}(2\ell)~\cong~\mathrm{U}(\ell)$, where $\mathrm{O}(\ell)$ and $\mathrm{U}(\ell)$ denote the orthogonal and unitary groups respectively. The equations above mean that any symplectic matrix can be (singular-value) decomposed into a symplectic-diagonal and PD matrix $\bm{D}$ together with two orthogonal-symplectic matrices $\bm{O}$ and $\bm{Q}$. This decomposition is known as the Bloch-Messiah, Euler or Symplectic SVD decomposition~\cite{serafini2017quantum}. 

To obtain this decomposition we first perform a polar decomposition on the input matrix
\begin{align}
\bm{S} = \bm{P} \bm{Y},
\end{align}
where $\bm{P}$ is PD and  symplectic and $\bm{Y}$ is orthogonal and symplectic, which implies that for some unitary matrix $\bm{U}$ of size $\ell$ we can write
\begin{align}
\bm{Y  }=\begin{pmatrix}
	\mathrm{Re}\left(\bm{U} \right) & -\mathrm{Im}\left(\bm{U} \right)\\
	\mathrm{Im}\left(\bm{U} \right) & \mathrm{Re}\left(\bm{U} \right)
\end{pmatrix},
\end{align}
as shown in Sec.~\ref{sec:symplectic}.

We now investigate the complex-version of the positive definite part,
\begin{align}
\bm{\mathcal{P}} = \bm{R}^\dagger \bm{P} \bm{R} = \begin{pmatrix}
	\bm{N} & \bm{M} \\
	\bm{M}^{*} & \bm{N}^{*}
\end{pmatrix},
\end{align}
where $\bm{M}=\bm{M}^{T}$ is symmetric, $\bm{N}$ is hermitian and PD as it is a diagonal block of a PD matrix. Since $\bm{\mathcal{P}}$ satisfies Eq.~\eqref{eq:complexsympform} we have that
\begin{align}
	\bm{N}^{2} &= \mathbb{1}_\ell + \bm{M}\bm{M}^{*}\nonumber\\
	&=\mathbb{1}_\ell +\bm{W}\bm{\Lambda}\bm{W}^{T}\bm{W}^{*}\bm{\Lambda}\bm{W}^{\dagger}\nonumber\\
	&=\bm{W}\bm{W}^{\dagger} +\bm{W}\bm{\Lambda}^{2}\bm{W}^{\dagger}\nonumber\\
	&=\bm{W}\left(\mathbb{1}_\ell+\bm{\Lambda}^{2}   \right)\bm{W}^{\dagger}\nonumber\\
	&=\bm{W}\sqrt{\mathbb{1}_\ell+\bm{\Lambda}^{2}}\sqrt{\mathbb{1}_\ell+\bm{\Lambda}^{2}}\bm{W}^{\dagger}\nonumber\\ &=\bm{W}\sqrt{\mathbb{1}_\ell+\bm{\Lambda}^{2}}\bm{W}^{\dagger}\bm{W}\sqrt{\mathbb{1}_\ell+\bm{\Lambda}^{2}}\bm{W}^{\dagger}\nonumber\\
	&=\left( \bm{W}\sqrt{\mathbb{1}+\bm{\Lambda}^{2}}\bm{W}^{\dagger} \right)^{2}.
\end{align}
Since $\bm{N}$ is PD, so is $\bm{N}^{2}$ and as such it has a unique square-root. Therefore, by knowing the Takagi/Autonne decomposition of $\bm{M}$ we can obtain the eigendecomposition of $\bm{N}$
\begin{align}
	\bm{N} = \bm{W}\sqrt{\mathbb{1}_\ell+\bm{\Lambda}^{2}}\bm{W}^{\dagger}.
\end{align}
Note that it is important to use the Takagi/Autonne decomposition of $\bm{M}$ and not the eigendecomposition of $\bm{N}$. This is because the latter in general does not give complete information about the unitary we are after. Indeed, consider the case where $\bm{\Lambda} = x \mathbb{1}_\ell$, then $\bm{N} = \sqrt{1+x^2} \mathbb{1}_{\ell}$ (from which we cannot obtain $\bm{W}$). On the other hand $\bm{M} = x \bm{W}\bm{W}^T$ which does contain information about $\bm{W}$.

We can combine our findings on $\bm{N}$ and $\bm{M}$ to write 
\begin{align}\label{eq:P}
\bm{\mathcal{P}} =& \begin{pmatrix}
	\bm{W} & 0_{\ell} \\
	0_{\ell} & \bm{W}^{*}
\end{pmatrix}\begin{pmatrix}
	\sqrt{\mathbb{1}_\ell+\bm{\Lambda}^{2}}& \bm{\Lambda} \\
	\bm{\Lambda} & \sqrt{\mathbb{1}+\bm{\Lambda}^{2}}
\end{pmatrix}\begin{pmatrix}
	\bm{W}^{\dagger} & 0_{\ell} \\
	0_{\ell} & \bm{W}^{T}
\end{pmatrix}\\
=& \exp\left[ \begin{pmatrix}
	0_\ell & \bm{W} \sinh^{-1} (\bm{\Lambda}) \bm{W}^T \\
	\bm{W}^* \sinh^{-1}( \bm{\Lambda}) \bm{W}^\dagger & 0_\ell
\end{pmatrix} \right].
\end{align}
In the last equation we used the fact that $\exp\left[\left(\begin{smallmatrix}
    0 & x \\
    x & 0
\end{smallmatrix} \right) \right] = \left(\begin{smallmatrix}
    \cosh x & \sinh x \\
    \sinh x & \cosh x
\end{smallmatrix} \right)$.
Combining Eq.~\eqref{eq:P} with Eq.~\eqref{eq:Ublock} and Eq.~\eqref{eq:complexpolar}, we find that the Bloch-Messiah/Euler decomposition in complex form is
\begin{align}
    \bm{\mathcal{S}} =& \begin{pmatrix}
	\bm{W} & 0_{\ell} \\
	0_{\ell} & \bm{W}^{*}
\end{pmatrix}\begin{pmatrix}
	\sqrt{\mathbb{1}+\bm{\Lambda}^{2}}& \bm{\Lambda} \\
	\bm{\Lambda} & \sqrt{\mathbb{1}+\bm{\Lambda}^{2}}
\end{pmatrix}\begin{pmatrix}
	\bm{X} & 0_{\ell} \\
	0_{\ell} & \bm{X}^{*}
\end{pmatrix}\\
=& \begin{pmatrix}
\bm{W} \sqrt{\mathbb{1}_\ell+\bm{\Lambda}^2} \bm{X} & \bm{W} \bm{\Lambda} \bm{X}^*\\
\bm{W}^* \bm{\Lambda} \bm{X}& \bm{W}^* \sqrt{\mathbb{1}_\ell+\bm{\Lambda}^2} \bm{X}^*
\end{pmatrix}
\end{align}
with $\bm{X} = \bm{W}^\dagger \bm{U}$.

In real form we have that
\begin{align}\label{eq:eigen-symplectic}
\bm{P} &= \bm{O} \bm{D} \bm{O}^T,\\
\bm{O} &= \begin{pmatrix}
	\mathrm{Re}\left(\bm{W} \right) & -\mathrm{Im}\left(\bm{W} \right)\\
	\mathrm{Im}\left(\bm{W} \right) & \mathrm{Re}\left(\bm{W} \right)
\end{pmatrix},\\
\bm{D}&=\bm{\Gamma} \oplus \bm{\Gamma}^{-1} \text{ with } \bm{\Gamma} = \bm{\Lambda} + \sqrt{\mathbb{1}_\ell + \bm{\Lambda}^2}.
\end{align}
Note that $\bm{D}$ has the desired form (as it is diagonal and symplectic) and that $\bm{O}\in \mathrm{C}(\ell)$. We can then define $\bm{Q} = \bm{O}^T \bm{Y}$ and complete the decomposition. Note that the multiplicative inverse of $\bm{\Lambda} + \sqrt{\mathbb{1}_\ell + \bm{\Lambda}^2}$ is precisely $-\bm{\Lambda} + \sqrt{\mathbb{1}_\ell + \bm{\Lambda}^2}$.

If we partition the real PD matrix as
\begin{align}
\bm{P} = \left( \begin{array}{cc}
	\bm{A} & \bm{B} \\
	\bm{B}^T & \bm{C}
\end{array}\right),
\end{align}
then we can obtain
\begin{align}
\bm{M} = \tfrac12\left( \bm{A} -\bm{C} + i \left[ \bm{B}+\bm{B}^T\right] \right),
\end{align}
whose Takagi/Autonne decomposition gives directly $\bm{D}$ and $\bm{O}$ in Eq.~\eqref{eq:eigen-symplectic}. The decomposition is summarized in the box below.

\begin{Frame}[Box 2. Bloch-Messiah/Euler decomposition]
	Input: Real symplectic matrix $\bm{S}$,  $\bm{S\Omega S}^T = \bm{\Omega}$.\\
	Obtain polar decomposition $\bm{S} = \bm{P} \bm{Y}$.\\
	Partition symplectic $\bm{P} = \left( \begin{array}{cc}
		\bm{A} & \bm{B} \\
		\bm{B}^T & \bm{C}
	\end{array}\right) $ .\\
	Obtain $\bm{M} =\tfrac12\left( \bm{A} -\bm{C} + i \left[ \bm{B}+\bm{B}^T\right] \right)$.\\
	Takagi/Autonne decompose $\bm{M} = \bm{W} \bm{\Lambda} \bm{W}^T.$ \\
	Form 
	$\bm{O} = \begin{pmatrix}
		\mathrm{Re}\left(\bm{W} \right) & -\mathrm{Im}\left(\bm{W} \right)\\
		\mathrm{Im}\left(\bm{W} \right) & \mathrm{Re}\left(\bm{W} \right)
	\end{pmatrix}$, \\
$\bm{\Gamma} =  \bm{\Lambda} + \sqrt{\mathbb{1}_\ell + \bm{\Lambda}^2}$, 
	$\bm{D}=\bm{\Gamma} \oplus \bm{\Gamma}^{-1}$.\\
	Obtain $\bm{Q} = \bm{O}^T \bm{Y}$.\\
	Return $\bm{O}, \bm{D}, \bm{Q}$.
\end{Frame}
Note that this decomposition is unique up to permutations and or degeneracies of the Takagi/Autonne singular values.

\section{Pre-Iwasawa and Iwasawa Decompositions}\label{sec:iwasawa}
The Iwasawa decomposition gives the decomposition of a symplectic matrix in terms of three matrices from different subgroups of the symplectic group~\cite{arvind1995real}. Two of these groups have already appeared before, namely the group of diagonal symplectic matrices (which is Abelian) and the compact subgroup $\mathrm{C}(\ell)$ which is isomorphic to the unitary group $\mathrm{U}(\ell)$. The third subgroup that we will need for the Iwasawa decomposition is the nilpotent symplectic subgroup $\mathrm{N}(\ell)$. The matrices from this subgroup have the form
\begin{align}\label{eq:nil}
\begin{pmatrix}
\bm{A} & 0_\ell \\
\bm{C} & \left[ \bm{A}^{-1} \right]^T
\end{pmatrix} \in \mathrm{N}(\ell)
\end{align}
where the square blocks satisfy $\bm{A}^T \bm{C} = \bm{C} \bm{A}^T$ and moreover $\bm{A}$ is lower unit triangular, i.e., $A_{i,i}=1, A_{i,j} = 0$ if $i<j$. It is direct to verify that matrices of the form in Eq.~\eqref{eq:nil} are symplectic and form a matrix group (i.e., they are a closed set under matrix multiplication and the inverses belong in the same set). 

Before obtaining the Iwasawa factorization, we compute the pre-Iwasawa factorization, following Arvind~\cite{arvind1995real}. To obtain the decomposition we first partition the to-be-factorized symplectic matrix as in Eq.~\eqref{eq:sympblocks} and define
\begin{align}
\bm{A}_0 &= \sqrt{\bm{A} \bm{A}^T + \bm{B} \bm{B}^T}, \\
\bm{C}_0 &= \left( \bm{C}\bm{A}^T+\bm{D}\bm{B}^T\right) \bm{A}_0^{-1},\\
\bm{X} &= \bm{A_0}^{-1} \bm{A},\\
\bm{Y} &= \bm{A_0}^{-1} \bm{B}.
\end{align}
We now prove some useful properties of these quantities.
First note that $\bm{A}_0^2$ is the top left block of the product $\bm{S} \bm{S}^T$. Since this product is PD, then so is $\bm{A}_0 = \bm{A}_0^T$. 
Second we note that the product $\bm{C}_0 \bm{A}_0^{-1} = \bm{A}_0^T \bm{C}_0^T$ is symmetric. To this end we simply note that
\begin{align}
\bm{C}_0 \bm{A}_0^{-1} = \bm{A}_0^{-1} \bm{C}_0^T \longleftrightarrow \bm{C}_0 = \bm{A}_{0}^{-1} \bm{C}_0^T \bm{A}_0.
\end{align}
The equality on the right-hand side is derived by using the properties of the blocks (cf. Eq.~\eqref{eq:sympblocks}) as follows
\begin{align}
\bm{C}_0  &= \bm{A}_0^{-2}\bm{A}_0^2\bm{C}_0\\
&= \bm{A}_0^{-2}\bigl(\bm{A}\bm{A}^T + \bm{B}\bm{B}^T\bigr)\bigl( \bm{C}\bm{A}^T+\bm{D}\bm{B}^T\bigr)\bm{A}_0^{-1} \\
&= \bm{A}_0^{-2}\bigl(\bm{A}\bm{A}^T\bm{C}\bm{A}^T +\bm{A}\bm{A}^T\bm{D}\bm{B}^T\nonumber\\
& \qquad + \bm{B}\bm{B}^T\bm{C}\bm{A}^T+ \bm{B}\bm{B}^T\bm{D}\bm{B}^T\bigr) \bm{A}_0^{-1}\\
\label{eqs:preiwasawasymidentities}&= \bm{A}_0^{-2}\bigl(\bm{A}\bm{C}^T\bm{A}\bm{A}^T +\bm{A}(\mathbb{1}_\ell+\bm{C}^T\bm{B})\bm{B}^T\nonumber\\
& \qquad + \bm{B}(\bm{D}^T\bm{A}-\mathbb{1}_\ell)\bm{A}^T+ \bm{B}\bm{D}^T\bm{B}\bm{B}^T\bigr)\bm{A}_0^{-1} \\
&= \bm{A}_0^{-2}\bigl(\bm{A}\bm{C}^T+\bm{B}\bm{D}^T \bigr)\bigl(\bm{A}\bm{A}^T +\bm{B}\bm{B}^T\bigr)\bm{A}_0^{-1} \\
&= \bm{A}_0^{-1} \bm{C}_0^T \bm{A}_0.
\end{align}
Finally, we now show that $\bm{U} = \bm{X}+ i \bm{Y}$ is unitary.
To this end consider the complex matrix $\bm{A} + i \bm{B}$.
If we do a polar decomposition of this matrix we easily find the square of the PD part to be
\begin{align}
 (\bm{A} + i \bm{B}) (\bm{A} + i \bm{B})^\dagger  = \underbrace{\bm{A} \bm{A}^T + \bm{B} \bm{B}^T}_{\equiv \bm{A}_0^2} + i (\underbrace{\bm{B} \bm{A}^T - \bm{A}^T \bm{B}}_{\equiv 0  \text{, cf. Eq.~\eqref{eq:sympblocks}}}).
\end{align}
Having the PD part, the unitary part is simply $\bm{A}_0^{-1} (\bm{A} + i \bm{B})$, completing the proof. 

With these properties proven we now state the (unique) pre-Iwasawa decomposition of a real symplectic matrix as
\begin{align}
\bm{S}  &=\bm{E} \bm{D}\bm{F},
\end{align}
where
\begin{align}
 \bm{E}=\begin{pmatrix}
		\mathbb{1}_\ell & 0_\ell\\
		\bm{C}_0 \bm{A}_0^{-1} & \mathbb{1}_\ell
	\end{pmatrix},
  \bm{D}=\begin{pmatrix}
		\bm{A}_0 & 0_\ell \\
		0_\ell & \bm{A}_0^{-1}
	\end{pmatrix},
 \bm{F}=\begin{pmatrix}
		\bm{X} & \bm{Y}\\
		-\bm{Y}  & \bm{X}
	\end{pmatrix},
\end{align}
such that $\bm{A}_0=\bm{A}_0^T$ and $\bm{F}\in \mathrm{C}(\ell)$. It can be straightforwardly computed as indicated in Box 3. One can verify that the decomposition is correct by multiplying the matrices in the equation above and using the properties of the blocks of a symplectic matrix (cf. Eq. ~\eqref{eq:sympblocks}). These details are presented in Appendix~\ref{app:preiwasawa}.
Note that matrices of the form $\bm{E}$ as in the last equation form an Abelian subgroup of the symplectic group. This group is isomorphic to the group of real symmetric matrices under matrix addition.
Note also that matrices of the form $\bm{D}$ are not a subgroup under matrix multiplication. The Iwasawa decompositon further decomposes the set of matrices above to obtain a unique decomposition in which each element of the decomposition belongs to a subgroup.

\begin{Frame}[Box 3. Pre-Iwasawa decomposition]
  Input: Real symplectic matrix $\bm{S}$,  $\bm{S\Omega S}^T = \bm{\Omega}$.\\
  Partition symplectic $\bm{S} = \left( \begin{array}{cc}
		\bm{A} & \bm{B} \\
		\bm{C} & \bm{D}
	\end{array}\right) $ .\\
	Obtain $\bm{A_0} = \sqrt{\bm{A}\bm{A}^T + \bm{B}\bm{B}^T}$ and  \\ $\bm{X}=\bm{A}_0^{-1}\bm{A}$,
    $\bm{Y}=\bm{A}_0^{-1}\bm{B}$,\\
    $\bm{C}_0 = \left( \bm{C}\bm{A}^T+\bm{D}\bm{B}^T\right) \bm{A}_0^{-1}$.\\
	Form 
	$\bm{E} = \begin{pmatrix}
		\mathbb{1}_\ell & 0_\ell\\
		\bm{C}_0 \bm{A}_0^{-1} & \mathbb{1}_\ell
	\end{pmatrix}$, \\
	$\bm{D} = \begin{pmatrix}
		\bm{A}_0 & 0_\ell \\
		0_\ell & \bm{A}_0^{-1}
	\end{pmatrix}$, \\	
    $\bm{F} = \begin{pmatrix}
		\bm{X} & \bm{Y}\\
		-\bm{Y}  & \bm{X}
	\end{pmatrix}$. \\
	Return $\bm{E}$, $\bm{D}$, $\bm{F}$.
\end{Frame}

To obtain the Iwasawa decomposition we follow Benzi and Razouk ~\cite{benzi2007iwasawa} and use the QR decomposition of the blocks in the central term of the pre-Iwasawa decomposition.
\begin{align}
\bm{A_0} = \bm{A_0}^T = \bm{Q} \bm{R}  = \bm{R}^T \bm{Q}^T
\end{align}
and writing 
\begin{align}
\bm{R}^T = \bm{R}^T \left( \bm{D}_a \bm{D}_s \right)^{-1} \bm{D}_a \bm{D}_s 
\end{align}
where $\bm{D}_a = \oplus_{i=1}^\ell |R_{i,i}|$ and $\bm{D}_s = \oplus_{i=1}^\ell \sign(R_{i,i})$. The matrix $\tilde{\bm{R}}^T = \bm{R}^T \left( \bm{D}_a \bm{D}_s \right)^{-1}$ is lower unit triangular and the matrix  $\tilde{\bm{Q}}^T = \bm{D}_s \bm{Q}^T$ is orthogonal. This allows us to write
\begin{align}
\bm{A}_0 &= \tilde{\bm{R}}^T \bm{D}_a \tilde{\bm{Q}}^T \\
\bm{A}_0^{-1} &= [\bm{A}_0^T]^{-1} = \tilde{\bm{R}}^{-1} \bm{D}_a^{-1} \tilde{\bm{Q}}^T 
\end{align} 
With this observation we can write 
\begin{align}
\bm{E} \bm{D} \bm{F} &= \bm{E} \begin{pmatrix}
	\bm{A}_0 & 0_\ell \\ 0_\ell & \bm{A}_0^{-1}
	\end{pmatrix}
	 \bm{F} \\
&=	 \underbrace{\bm{E} \begin{pmatrix}
	 	\tilde{\bm{R}}^T & 0_\ell \\ 0_\ell & \tilde{\bm{R}}^{-1}
	 \end{pmatrix}}_{\equiv \tilde{\bm{E}}} \underbrace{ \begin{pmatrix}
	 	\bm{D}_a & 0_\ell \\ 0_\ell & \bm{D}_a^{-1}
	 \end{pmatrix} }_{\equiv \tilde{\bm{D}}} \underbrace{ \begin{pmatrix}
	 \bm{D}_s{\bm{Q}^T} & 0_\ell \\ 0_\ell & \bm{D}_s{\bm{Q}^T}
 \end{pmatrix} \bm{F}}_{\equiv \tilde{\bm{F}}}.
\end{align}
Note that $\tilde{\bm{F}}$ is part of the compact subgroup as it is the product of two elements from this set. It is also direct to verify that the matrix $\tilde{\bm{E}}$ has precisely the form in Eq.~\eqref{eq:nil}. 

\begin{Frame}[Box 4. Iwasawa decomposition]
  Input: Real symplectic matrix $\bm{S}$,  $\bm{S\Omega S}^T = \bm{\Omega}$.\\
  Pre-Iwasawa decompose to obtain $\bm{E}$, $\bm{D}$, $\bm{F}$ \, .\\ 
  Identify elements $\bm{D} = \begin{pmatrix}
		\bm{A}_0 & 0_\ell \\
		0_\ell & \bm{A}_0^{-1}
	\end{pmatrix}$, \\	
  Obtain QR-decomposition $\bm{A}_0=\bm{Q}\bm{R} = \bm{R}^T\bm{Q}^T$ \,.\\  
  Form diagonal matrices $\bm{D}_a =\oplus_{i=1}^\ell |R_{i,i}| $ and $\bm{D}_s = \oplus_{i=1}^\ell \sign( R_{i,i})$\\
    Form $\tilde{\bm{R}}= \bm{D}_s\bm{D}_a\bm{R}$\, \\
       $ \tilde{\bm{E}}= \bm{E}\begin{pmatrix}
		\tilde{\bm{R}}^T & 0_\ell\\
		0_\ell & \tilde{\bm{R}}^{-1}
	\end{pmatrix}$ \, ,\\
 	$\tilde{\bm{D}} = \begin{pmatrix}
		\bm{D}_a & 0_\ell \\
		0_\ell & \bm{D}_a^{-1}
	\end{pmatrix}$ \,, \\	
     $\tilde{\bm{F}} = \begin{pmatrix}
		\bm{D}_s\bm{Q}^T & 0_\ell\\
		0_\ell & \bm{D}_s\bm{Q}^T
	\end{pmatrix} \bm{F}$ \\
	Return $\tilde{\bm{E}}$, $\tilde{\bm{D}}$, $\tilde{\bm{F}}$.
\end{Frame}

\section{Williamson Decomposition}\label{sec:will}

Williamson's theorem states that given a real symmetric PD matrix $\bm{\Sigma} \in \mathbb{R}^{2\ell\times 2\ell}$ ($\bm{\Sigma}^{T} = \bm{\Sigma}>0$) there exists a real symplectic matrix $\bm{S}\in\mathrm{Sp}(2\ell,\mathbb{R})$ such that
\begin{align}
	\bm{\Sigma} = \bm{S}\bm{T}\bm{S}^{T},
\end{align}
where
\begin{align}
	\bm{T}=\left( \bm{\Delta}\oplus\bm{\Delta}\right), \quad \bm{\Delta}=\oplus_{i=1}^{\ell}\delta_{i},
\end{align}
with $\delta_i >  0$. Each value, $\delta_{i}$, is such that
\begin{align}\label{symp:evals}
	\det\left(\delta_{i}\mathbb{1}_{2\ell} + i\bm{\Omega}\bm{\Sigma}   \right)=0,
\end{align}
where $\bm{\Omega}$ is the symplectic form(recall Eq.~\eqref{eq:sympform}) \cite{williamson1936algebraic,nicacio2021williamson}. The values $\delta_{i}$ are thus the eigenvalues of the matrix $i \bm{\Omega}  \bm{\Sigma}$.

We follow Idel et al.\cite{idel2016operational,idel2017perturbation}
and define 
\begin{align}\label{eq:Mtilde}
\bm{\Psi} \equiv \left[ \sqrt{\bm{\Sigma}} \right] ^{-1} \bm{\Omega} \left[ \sqrt{\bm{\Sigma}} \right] ^{-1}.
\end{align} 
By construction,  $\bm{\Psi}$ is real and antisymmetric and so its real Schur decomposition (recall Sec.~\ref{sec:schur}) takes the form
\begin{align}
	\bm{\Psi} = \tilde{\bm{O}} [\oplus_{i=1}^{\ell}  \left(\begin{smallmatrix} 0 &\pm {\phi_i}  \\ \mp{\phi_i} & 0 \end{smallmatrix} \right)] \tilde{\bm{O}}^T,
\end{align}
where $\phi_i \neq 0 \ \forall i$, as $|\det(\bm{\Psi})|=\prod_{i=1}^\ell \phi_i^2 = 1/|\det(\bm{\Sigma})| \neq 0$ and $\tilde{\bm{O}}$ is an orthogonal matrix.

We can always insert a permutation matrix, $\bm{\Pi}_{1}$, such that all the positive values are above the diagonal
\begin{align}
	\bm{\Pi}_1^T [\oplus_{i=1}^{\ell}  \left(\begin{smallmatrix} 0 &\pm {\phi_i}  \\ \mp{\phi_i} & 0 \end{smallmatrix} \right)]  \bm{\Pi}_1 =  [\oplus_{i=1}^{\ell}  \left(\begin{smallmatrix} 0 &{\phi_i}  \\- {\phi_i} & 0 \end{smallmatrix} \right)],
\end{align}
which allows us to write
\begin{align}\label{eq:F}
	\bm{\Psi} = \tilde{\bm{O}} \bm{\Pi}_1 [\oplus_{i=1}^\ell  \left(\begin{smallmatrix} 0 & {\phi_i}  \\ -{\phi_i} & 0 \end{smallmatrix} \right)] \bm{\Pi}^T_1 \tilde{\bm{O}}^T.
\end{align}
Although this permutation depends on the output of the Schur decomposition algorithm used, we can express it in general as
\begin{align}
	\bm{\Pi}_{1}=\oplus_{i=1}^{\ell}\bm{w}_{i},
\end{align}
where $\bm{w}_{i}=\mathbb{1}_{2}$ if the positive value is above the diagonal and $\bm{w}_{i}=\left(\begin{smallmatrix} 0 & 1\\ 1 & 0 \\\end{smallmatrix}\right)$ if it is not.

Furthermore, we can always insert another permutation matrix, $\bm{\Pi}_{2}$, which transforms the block-diagonal matrix into an off-diagonal block matrix
\begin{align}\label{eq:permute2}
	\bm{\Pi}_2^T  \left[  \oplus_{i=1}^{\ell}  \left(\begin{smallmatrix} 0 & {\phi_i}  \\ -{\phi_i} & 0 \end{smallmatrix} \right) \right]  \bm{\Pi}_2  = \begin{pmatrix} 0_{\ell} &     \oplus_{i=1}^{\ell}   {\phi_i} \\ - \oplus_{i=1}^{\ell}  {\phi_i} & 0_{\ell} \end{pmatrix}.
\end{align}
This permutation is independent of the Schur decomposition. Although it is slightly more convoluted to express in matrix form, this permutation is commonly used in quantum optics and represents a change of basis from the $(x_{1},p_{1},\ldots,x_{\ell},p_{\ell})$ basis to the $(x_{1},\ldots,x_{\ell},p_{1},\ldots,p_{\ell})$ basis.
Collecting all the permutations and orthogonal matrices we define a new orthogonal matrix
\begin{align}
\bm{O} = \tilde{\bm{O}}\bm{\Pi}_1 \bm{\Pi}_2,
\end{align}
which by construction satisfies
\begin{align}\label{eq:Ocons}
\bm{O} \sqrt{\bm{\Phi} \oplus \bm{\Phi} } \bm{\Omega} \sqrt{\bm{\Phi} \oplus \bm{\Phi} } \bm{O} ^T = \bm{\Psi},
\end{align}
where we introduced the diagonal matrix $\bm{\Phi} = \oplus_{i=1}^\ell \phi_i$ and factored
\begin{align}
	\begin{pmatrix} 0_{\ell} &     \oplus_{i=1}^{\ell}   {\phi_i} \\ - \oplus_{i=1}^{\ell}  {\phi_i} & 0_{\ell} \end{pmatrix} =& \sqrt{\bm{\Phi} \oplus \bm{\Phi} } \bm{\Omega} \sqrt{\bm{\Phi} \oplus \bm{\Phi} } \\=&  [\bm{\Phi} \oplus \bm{\Phi}]  \bm{\Omega}.
\end{align}
We can equate \eqref{eq:Ocons} and \eqref{eq:F} and pre- and post- multiply both sides by $\sqrt{\bm{\Sigma}}$ to obtain
\begin{align}\label{eq:almostS}
\left[\sqrt{\bm{\Sigma}} \bm{O} \sqrt{\bm{\Phi} \oplus \bm{\Phi} } \right] \bm{\Omega} \left[ \sqrt{\bm{\Phi} \oplus \bm{\Phi} } \bm{O} ^T \sqrt{\bm{\Sigma}} \right]  = \bm{\Omega}.
\end{align}

Notice that the orthogonal matrix $\bm{O}$ could also be obtained by finding the eigendecomposition of the symmetric matrix~\cite{nicacio2021williamson}
\begin{align}
\bm{\Psi}^{-2} =  \sqrt{\bm{\Sigma}} \bm{\Omega} \bm{\Sigma}   \bm{\Omega} \sqrt{\bm{\Sigma}} = -\bm{O} (\bm{\Phi} \oplus \bm{\Phi} )^{-2} \bm{O}^T.
\end{align}
Numerically, directly obtaining the eigendecomposition of $\bm{\Psi}^{-2}$ does not always lead to the correct orthogonal matrix satisfying Eq.~\eqref{eq:almostS}, as each eigenvalue of $\bm{\Psi}^{-2}$ is pair-wise degenerate. In particular they can always be swapped which can lead to an incorrect symplectic form on the right-hand side of Eq.~\eqref{eq:almostS}, where not all the elements above the diagonal are non-negative.

Returning to our derivation, from Eq.~\eqref{eq:almostS}
we identify the symplectic matrix
\begin{align}
	\bm{S} = \sqrt{\bm{\Sigma}} \bm{O} \sqrt{\bm{\Phi} \oplus \bm{\Phi} } = \sqrt{\bm{\Sigma}} \tilde{\bm{O}}\bm{\Pi}_1 \bm{\Pi}_2\sqrt{\bm{\Phi} \oplus \bm{\Phi} }.
	\end{align}
Furthermore if we define $\bm{\Delta} = \oplus_{i=1}^\ell \delta_i =  \bm{\Phi}^{-1} = \oplus_{i=1}^\ell \phi_i^{-1}$, it directly follows that
\begin{align}
	\bm{S} [\bm{\Delta} \oplus \bm{\Delta}] \bm{S}^T = \bm{S} \bm{T} \bm{S}^T =  \bm{\Sigma},
\end{align}
completing the proof. 

 The derivation is summarized in the box below.
\begin{Frame}[Box 5. Williamson decomposition]
    Input: Real symmetric positive definite matrix $\bm{\Sigma}$, $\bm{\Sigma} = \bm{\Sigma}^{T}$, $\bm{\Sigma} \in \mathbb{R}^{2\ell\times 2\ell}$. \\
	Construct antisymmetric $\bm{\Psi} = \left[ \sqrt{\bm{\Sigma}} \right] ^{-1} \bm{\Omega} \left[ \sqrt{\bm{\Sigma}} \right] ^{-1}$.\\
    Calculate Schur decomposition\\  $\bm{\Psi} = \tilde{\bm{O}} [\oplus_{i=1}^{\ell}  \left(\begin{smallmatrix} 0 &\pm {\phi_i}  \\ \mp{\phi_i} & 0 \end{smallmatrix} \right)] \tilde{\bm{O}}^T$.\\
	Find permutation matrix $\bm{\Pi}_{1}$ given Schur output.\\
	Construct fixed permutation matrix $\bm{\Pi}_2$ which transforms $\bm{\Pi}_2 (x_{1},p_{1},\ldots,x_{\ell},p_{\ell})^T=(x_{1},\ldots,x_{\ell},p_{1},\ldots,p_{\ell})^T$.\\
	Construct $\bm{\Phi} = \oplus_{i=1}^{\ell} \phi_i $.\\
	Return: $\bm{S} = \sqrt{\bm{\Sigma}} \tilde{\bm{O}} \bm{\Pi}_1 \bm{\Pi}_2\left[\sqrt{\bm{\Phi} \oplus \bm{\Phi}}\right]$, \\ $\bm{T} = \bm{\Delta} \oplus \bm{\Delta} = \left[\bm{\Phi} \oplus \bm{\Phi} \right] ^{-1}$.
\end{Frame}

Note that the symplectic eigenvalues of $\bm{\Sigma}$ are identical to the absolute values of the ``regular'' eigenvalues of $ i \bm{\Omega} \bm{\Sigma}$. To this end, first recall Sylvester's determinant theorem~\cite{sylvester1883xxxix} (or Appendix B of \cite{pozrikidis2014introduction}), $\det(\mathbb{1}z - \bm{A} \bm{B}) = \det(\mathbb{1}z - \bm{B} \bm{A})$ for any two  matrices of compatible sizes $\bm{A}$ and $\bm{B}$. Consider now the characteristic polynomial of $i \bm{\Omega} \bm{\Sigma}$
\begin{align}
\det(\mathbb{1}_{2 \ell}z - i \bm{\Omega} \bm{\Sigma})=&\det(\mathbb{1}_{2 \ell}z - i \bm{\Omega} \bm{S} \bm{T} \bm{S}^T) \\
=&\det(\mathbb{1}_{2 \ell}z - i \bm{S}^T\bm{\Omega} \bm{S} \bm{T} ) \\
=&\det(\mathbb{1}_{2 \ell}z - i \bm{\Omega} \bm{T} )\\
=& \prod_{i=1}^\ell (z^2 - \delta_i^2),
\end{align}
as claimed in Eq.~\eqref{symp:evals}.

Note that the existence of this decomposition is discussed in e.g. theorem 1 of Ferraro, Olivares and Paris~\cite{ferraro2005gaussian} and that the symplectic matrix appearing in this decompositions can also be obtained by evaluating sub-matrix determinants~\cite{pereira2021symplectic}.

\section*{Acknowledgements}
    M.H. and N.Q. acknowledge support from the Minist\`{e}re de l'\'{E}conomie et de l’Innovation du Qu\`{e}bec and the Natural Sciences and Engineering Research Council of Canada. W.M. acknowledges support from European Research Council Starting grant (950402). M.H. and N.Q. thank J.E. Sipe for insightful discussions. N.Q. thanks S. Duque Mesa for comments on the \texttt{Python} implementation and H. de Guise for valuable discussions.

\appendix
\section{Proof of Pre-Iwasawa Decomposition}\label{app:preiwasawa}
The expressions for the pre-Iwasawa decomposition can be seen to hold true by simply multipying the purported decompositions and invoking the symplectic conditions Eq.~\eqref{eq:sympblocks} and Eq.~\eqref{eqs:preiwasawasymidentities} as follows:
\begin{align}
\bm{S} &= \left( \begin{array}{cc}
		\bm{A} & \bm{B} \\
		\bm{C} & \bm{D}
	\end{array}\right)\\
 &=\bm{E} \bm{D}\bm{F}\\
 &=\begin{pmatrix}
		\mathbb{1}_\ell & 0_\ell\\
		\bm{C}_0 \bm{A}_0^{-1} & \mathbb{1}_\ell
	\end{pmatrix}.\begin{pmatrix}
		\bm{A}_0 & 0_\ell \\
		0_\ell & \bm{A}_0^{-1}
	\end{pmatrix}. \begin{pmatrix}
		\bm{X} & \bm{Y}\\
		-\bm{Y}  & \bm{X}
	\end{pmatrix} \\
  &=\begin{pmatrix}
		\bm{A}_0 & 0_\ell \\
		\bm{C}_0 & \bm{A}_0^{-1}
	\end{pmatrix}. \begin{pmatrix}
		\bm{X} & \bm{Y}\\
		-\bm{Y}  & \bm{X}
	\end{pmatrix} \\
   &= \begin{pmatrix}
		\bm{A}_0\bm{X} & \bm{A}_0\bm{Y}\\
		\bm{C}_0\bm{X}-\bm{A}_0^{-1}\bm{Y}  &\bm{C}_0\bm{Y}+ \bm{A}_0^{-1}\bm{X}
	\end{pmatrix} 
\end{align}
To simplify the bottom blocks we do
\begin{align}
 \bm{C}_0\bm{X}&-\bm{A}_0^{-1}\bm{Y}\\
 &=\bm{C}_0\bm{A}_0^{-1}\bm{A}-\bm{A}_0^{-2}\bm{B}\\
 &=\bm{A}_0^{-1}\bm{C}_0^T\bm{A}-\bm{A}_0^{-2}\bm{B}\\
 &=\bm{A}_0^{-2}\biggl(\bigl(\bm{A}\bm{C}^T+\bm{B}\bm{D}^T \bigr)\bm{A}-\bm{B}\biggr)\\
 &=\bm{A}_0^{-2}\biggl(\bm{A}\bm{C}^T\bm{A}+\bm{B}\bm{D}^T \bm{A}-\bm{B}\biggr)\\
  &=\bm{A}_0^{-2}\biggl(\bm{A}\bm{A}^T\bm{C}+\bm{B}\bigl( \mathbb{1}_{\ell}+\bm{B}^T\bm{C}\bigr)-\bm{B}\biggr)\\
    &=\bm{A}_0^{-2}\bigl(\bm{A}\bm{A}^T+\bm{B}\bm{B}^T\bigr)\bm{C}= \bm{C},
\end{align}
\begin{align}
    \bm{C}_0\bm{Y}&+ \bm{A}_0^{-1}\bm{X}\\
 &=\bm{C}_0\bm{A}_0^{-1}\bm{B}+\bm{A}_0^{-2}\bm{A}\\
 &=\bm{A}_0^{-2}\biggl(\bigl(\bm{A}\bm{C}^T+\bm{B}\bm{D}^T \bigr)\bm{B}+\bm{A}\biggr)\\
  &=\bm{A}_0^{-2}\biggl(\bigl(\bm{A}\bm{C}^T+\bm{B}\bm{D}^T \bigr)\bm{B}+\bm{A}\biggr)\\
   &=\bm{A}_0^{-2}\biggl(\bm{A}\bm{C}^T\bm{B}+\bm{B}\bm{D}^T\bm{B}+\bm{A}\biggr)\\
   &=\bm{A}_0^{-2}\biggl(\bm{A}\bigl(\bm{A}^T\bm{C}-\mathbb{1}_{\ell}\bigr)+\bm{B}\bm{B}^T\bm{D}+\bm{A}\biggr)=\bm{D}.
\end{align}

\bibliography{bib.bib}
\end{document}